\documentclass[10pt, conference]{IEEEtran}
\usepackage{graphicx}
\usepackage{grffile}
\usepackage{framed}
\usepackage{amsmath}
\usepackage{booktabs,caption,fixltx2e}
\usepackage[flushleft]{threeparttable}
\usepackage{balance}
\usepackage[dvipsnames]{xcolor}
\usepackage{fancybox}
\usepackage{comment}
\usepackage{mathtools}
\usepackage{pdfpages}
\usepackage{lastpage}
\usepackage{fancyhdr}

\newcommand{\gopi}[1]{\textcolor{blue}{{\it#1}}}

\newcommand{\rqbox}[1]{
\begin{center}
\cornersize{.2}
\setlength{\fboxsep}{7pt}
\ovalbox{\vspace{-0.0cm}\begin{minipage}{3.1in}
{\em #1}
\end{minipage}}
\end{center}}

\usepackage{array}

\newcolumntype{R}[1]{>{\raggedleft\let\newline\\\arraybackslash\hspace{2pt}}m{#1}}

\ifCLASSINFOpdf
\else
\fi
\begin{document}
%
\title{The Impact of Using Regression Models to Build Defect Classifiers}

\author{\IEEEauthorblockN{Gopi Krishnan Rajbahadur, Shaowei Wang}
	\IEEEauthorblockA{Queen's University, Canada\\	\{krishnan, shaowei\}@cs.queensu.ca}
	\and
	\IEEEauthorblockN{Yasutaka Kamei}
	\IEEEauthorblockA{Kyushu University, Japan\\	kamei@ait.kyushu-u.ac.jp}
	\and
		\IEEEauthorblockN{Ahmed E. Hassan}
	\IEEEauthorblockA{Queen's University, Canada\\	ahmed@cs.queensu.ca}
}


\maketitle

\begin{abstract}

It is common practice to discretize continuous defect counts into defective and non-defective classes and use them as a target variable when building defect classifiers (discretized classifiers). However, this discretization of continuous defect counts leads to information loss that might affect the performance and interpretation of defect classifiers. Another possible approach to build defect classifiers is through the use of regression models then discretizing the predicted defect counts into defective and non-defective classes (regression-based classifiers).

In this paper, we compare the performance and interpretation of defect classifiers that are built using both approaches (i.e., discretized classifiers and regression-based classifiers) across six commonly used machine learning classifiers (i.e., linear/logistic regression, random forest, KNN, SVM, CART, and neural networks) and 17 datasets. We find that: i) Random forest based classifiers outperform other classifiers (best AUC) for both classifier building approaches; ii) In contrast to common practice, building a defect classifier using discretized defect counts (i.e., discretized classifiers) does not always lead to better performance.

Hence we suggest that future defect classification studies should consider building regression-based classifiers (in particular when the defective ratio of the modeled dataset is low). Moreover, we suggest that both approaches for building defect classifiers should be explored, so the best-performing classifier can be used when determining the most influential features.



\end{abstract}


%
\IEEEpeerreviewmaketitle

\section{Introduction}
\label{sec:Introduction}
Finding and fixing defects consume more than 80\% of the total budget of a software project. These costs can be reduced significantly if the defects are identified and fixed early on~\cite{3arar2015software,4ceylan2006software,5fagan2001design,6moser2008comparative,7mullen2005software,8shull2002we}.

Defect classifiers assist in software quality assurance efforts and in prioritizing process improvement efforts. In particular, defect classifiers can identify defect-prone modules~\cite{11hassan2009predicting,10kim2007predicting,wang2013using,51zimmermann2007predicting}, in turn helping quality assurance teams allocate their limited resources to these modules (e.g., packages, files, or classes). Moreover, the trained defect classifiers can be used to understand the impact of the various features (e.g., process or product metrics) on the defect-proneness of a module, in turn helping practitioners (through process improvement efforts) avoid pitfalls that have led to defective modules in the past.


The most common approach to build a classifier is through the discretization of the continuous defect counts into ``defective'' and ``non-defective'' classes and using these classes as a target variable (i.e., \emph{discretized defect classifiers} )~\cite{12cataldo2009software,28lessmann2008benchmarking,14mockus2010organizational}. However, the discretization of continuous variables (i.e., defect counts in this case) into two classes often leads to a significant loss of information~\cite{17altman2006cost,16cohen1983cost,18royston2006dichotomizing} and introduces undesired false positives or false negatives~\cite{19austin2004inflation}.

To avoid the information loss because of discretization, one possible solution is to perform classification via regression~\cite{23hou2013efficient,21Singh06,22xiang2010semi}. Classification via regression first builds a regression model using the non-discretized defect counts then uses the predicted defect counts to identify the presence or absence of defect (i.e., \emph{regression-based defect classifiers}.) 
However, it is not clear which classifier building approach leads to better performing defect classifiers. It is also not clear whether these two approaches would produce classifiers which are influenced by different set of features (e.g., product versus process metrics).



In this paper, we examine the use of regression models to build defect classifiers in terms of performance (i.e., Area Under the receiver operator characteristic Curve (AUC)) and model interpretation (i.e., influential features that impact the defect-proneness of a module). We conduct our study on six commonly used classifiers (i.e., linear/logistic regression, random forest, KNN, SVM, CART, and neural networks) and using 17 Tera-PROMISE defect datasets~\cite{25promiserepo}. We conduct our study through the following research questions:

\begin{itemize}
 \item \textbf{RQ1. How well do regression-based classifiers perform?}\\
        In contrast to current practices in our field, building classifiers using discretized defect counts does not always lead to better performance.  Regression-based classifiers outperform discretized classifiers when the defective ratio of the modeled dataset is low ($< 15\%$) and the pattern reverses when the defective ratio is high ($> 35\%$) for the random forest classifier.
  \item \textbf{RQ2. Are discretized and regression-based classifiers influenced by the same set of features?}\\
	    The most influential features (i.e., Rank 1 features) do not vary significantly between both approaches for building a defect classifier (when using a random forest based classifier). However, we observe significant variances for features at lower ranks.
	\end{itemize}

	Thus we suggest that future defect classification studies should consider building regression-based classifiers (in particular when the defective ratio of the modeled dataset is low). Moreover, we suggest that both approaches for building defect classifiers should be explored, so the best-performing classifier can be used when determining the most influential features.


\textbf{Paper organization\gopi{:}} Section~\ref{sec:Background} presents the background and related work. In section~\ref{sec:ExperimentSetup}, we describe the data collection and overall approach of our study and present a preliminary study in Section~\ref{sec:prestudy}, The results of our RQs and their implications are discussed in Section~\ref{sec:Results}. In Section~\ref{sec:Discussion}, we delve deeper into some of our findings and discuss the threats to validity in Section~\ref{sec:ThreatsToValidity}. Finally, we conclude our paper in Section~\ref{sec:Conclusion}.

\section{Background and Related Work}
\label{sec:Background}

\subsection{Defect Prediction}
Defect classifier can be used to identify modules that are likely to be defective (i.e., defect-prone)~\cite{10kim2007predicting,menzies2013local,nam2015clami,scanniello2013class,wang2013using,51zimmermann2007predicting}. Quality assurance teams can then effectively focus their limited testing and code review resources on such defect-prone modules. For instance,
Kim et al.~\cite{10kim2007predicting} propose an approach to predict defective modules by leveraging software change history data. Second, defect classifiers can be used to plan process improvement efforts by focusing on controlling the most influential features in the classifiers (i.e., model interpretation)~\cite{bavota2015impact,12cataldo2009software,di2015role,13mcintosh2014impact,14mockus2010organizational,Posnett:2013:DEM:2486788.2486848,Rahman:2013:WPM:2486788.2486846}. For example, Cataldo et al.~\cite{12cataldo2009software} employ a logistic classifier to identify that defects are influenced by workflow dependencies over syntactic dependencies. In this paper, we propose the use of regression models to build defect classifiers in lieu of the common use of machine learning classifiers. We structure our analysis in terms of performance (i.e., AUC) and model interpretation (i.e., influential features that impact the defect-proneness of a module).

\subsection{Discretization of a Continuous Variable}
Discretization is referred as the process of transferring a continuous variable into its discrete classes. Prior studies point out that discretization increases the risk of information loss and introduction of false positives~\cite{17altman2006cost,19austin2004inflation,16cohen1983cost,20maccallum2002practice,18royston2006dichotomizing}, which is due to the loss of variance that happens to continuous target variable (defect counts) during discretization~\cite{16cohen1983cost}.

Traditionally, prior defect studies usually do not delve into such information loss. These studies discretize the defect counts into ``defective'' and ``non-defective'' classes then use the discretized defect counts to build a classifier. To alleviate the information loss problem, one solution is to build a classifier through regression, which uses the non-discretized defect counts to build a regression model first, then performs classification based on the results from the regression models~\cite{23hou2013efficient,21Singh06,22xiang2010semi}.

In classification via regression the continuous defect counts are used as a target variable directly and the resulting prediction is used for classification based on a threshold.  However such an approach is discouraged in the machine learning literature as it is much more sensitive to the fluctuation in data than traditional classification techniques~\cite{24ng2003lecture}.

In summary, both discretized and regression-based defect classifiers have inherent flaws and it is not clear which approach is better. In this paper, we compare the traditional classification approach against the classification via regression approach for building defect classifiers in order to better understand both approaches in the context of software defect data.

\section{Experimental Setting}
\label{sec:ExperimentSetup}
This section describes the data collection, and gives an overview of our study approach.

\begin{figure*}
    \includegraphics[width=\linewidth]{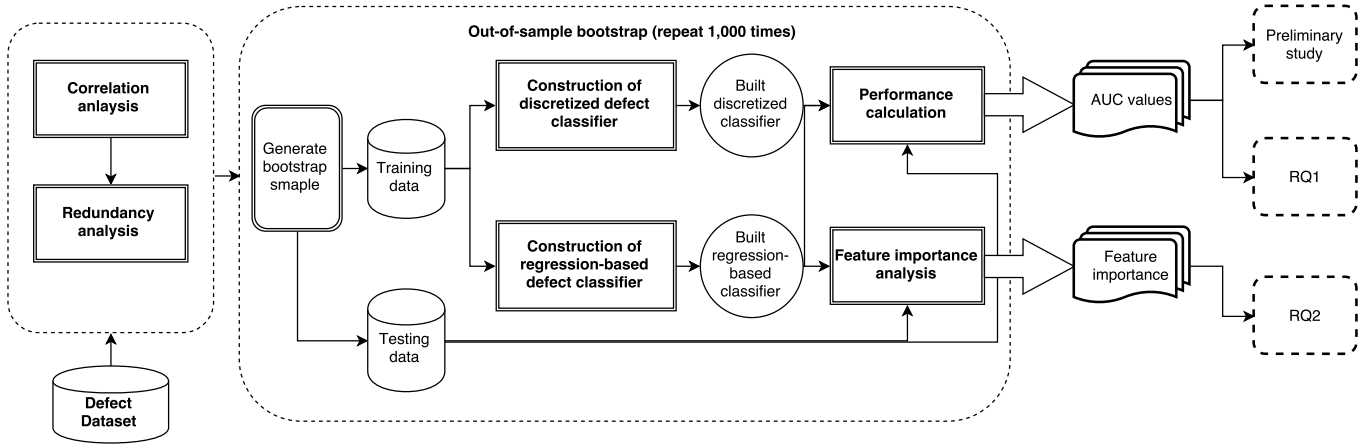}
    \caption{An overview of our study approach.}
    \label{fig:framework}
\end{figure*}
\begin{table}
	
	\caption{Overview of the studied datasets.}\label{tab:a}
	\begin{tabular}{l|p{1cm}rrp{1cm}r}
		\hline
		\textbf{Project} & \textbf{DR(\%)}  & \textbf{\#Files} & \textbf{\#Features} & \textbf{\#FACRA}& \textbf{EPV}\\
		\hline
		Eclipse-2.0 & \multicolumn{1}{r}{14.5} & 6,729 & 32 & \multicolumn{1}{r}{12} & 30\\
		Eclipse-2.1 & \multicolumn{1}{r}{10.8} & 7,888 & 32 & \multicolumn{1}{r}{12} & 30\\
		Eclipse-3.0 & \multicolumn{1}{r}{14.8} & 10,593 & 32 & \multicolumn{1}{r}{12} & 49\\
		Camel-1.2 & \multicolumn{1}{r}{35.5} & 608 & 20 & \multicolumn{1}{r}{12} & 11\\
		Mylyn & \multicolumn{1}{r}{13.2} & 1,862 & 15 & \multicolumn{1}{r}{8} & 16\\
		PDE & \multicolumn{1}{r}{14.0} & 1,497 & 15 & \multicolumn{1}{r}{9} & 14\\
		Prop-1 & \multicolumn{1}{r}{14.8} & 18,471 & 20 & \multicolumn{1}{r}{15} & 137\\
		Prop-2 & \multicolumn{1}{r}{10.6} & 23,014 & 20 & \multicolumn{1}{r}{14} & 122\\
		Prop-3 & \multicolumn{1}{r}{11.5} & 10,274 & 20 & \multicolumn{1}{r}{15} & 59\\
		Prop-4 & \multicolumn{1}{r}{9.6} & 8,718 & 20 & \multicolumn{1}{r}{15} & 42\\
		Prop-5 & \multicolumn{1}{r}{15.3} & 8,516 & 20 & \multicolumn{1}{r}{14} & 65\\
		Xalan-2.5 & \multicolumn{1}{r}{48.2} & 803 & 20 & \multicolumn{1}{r}{14} & 19\\
		Xalan-2.6 & \multicolumn{1}{r}{46.4} & 885 & 20 & \multicolumn{1}{r}{13} & 21\\
		Lucene-2.4 & \multicolumn{1}{r}{59.7} & 340 & 20 & \multicolumn{1}{r}{13} & 10\\
		Poi-2.5 & \multicolumn{1}{r}{64.4} & 385 & 20 & \multicolumn{1}{r}{12} & 12\\
		Poi-3.0 & \multicolumn{1}{r}{63.6} & 442 & 20 & \multicolumn{1}{r}{13	} & 14\\
		Xerces-1.4 & \multicolumn{1}{r}{74.3} & 588 & 20 & \multicolumn{1}{r}{11} & 22\\

		\hline
	\end{tabular}
	\begin{tablenotes}
		\scriptsize
		\item DR - Defective Ratio; FACRA - Features After Correlation and Redundancy Analysis
	\end{tablenotes}
	
\end{table}
\begin{figure}
	\includegraphics[width=\linewidth]{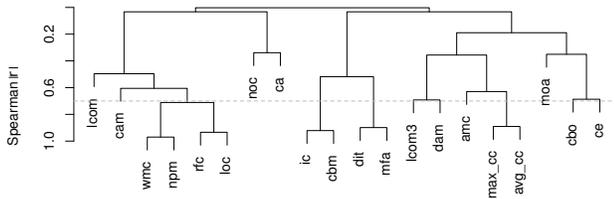}
	\vspace{-2cm}
	\caption{Clustered features after correlation analysis for Poi-3.0. The dotted line represents the 0.7 threshold and we select only one feature from the sub-hierarchy below for inclusion in our analysis. }
	
	\label{fig:exp}
\end{figure}

\subsection{Data collection}

We use data from the Tera-PROMISE Repository~\cite{25promiserepo}. Tera-PROMISE contains 101 software projects data, and the types of these projects are diverse. Using data from Tera-PROMISE helps us draw more general observations across different datasets. We select datasets based on the following two criteria which are similar to a prior study~\cite{37tantithamthavorn2016automated}:

\textbf{Criterion 1: Remove datasets with an EPV that is larger than 10.}
Events Per Variable (EPV) is defined as the ratio of the frequency of the least occurring class in the outcome variable to the number of features that are involved in training of a classifier.
Prior studies show that the EPV value has a significant influence on the performance of defect classifiers~\cite{39peduzzi1996simulation,40tantithamthavorn2016empirical}. In particular,  defect classifiers trained with datasets with a low EPV value yield unstable results~\cite{37tantithamthavorn2016automated,40tantithamthavorn2016empirical}. To ensure the stability of our results, we select datasets with an EPV value that is larger than 10~\cite{39peduzzi1996simulation}. We calculate the EPV for our datasets using the steps that are provided by Tantithamthavorn et al.~\cite{40tantithamthavorn2016empirical}.

\textbf{Criterion 2: Remove datasets that have more than 80\% defective modules.} We choose datasets that have less than 80\% defective modules, because it is highly unlikely for any software project to have modules with defects that are considerably more than clean modules.

Among the 101 Tera-PROMISE datasets, we excluded 78 datasets since they had an EPV value that is less than 10. To satisfy criterion 2, we eliminate the Xalan-2.7 project and end up with 22 datasets that satisfy our criteria. We had to eliminate another 5 datasets as they did not have the actual defect counts for each module (they only had the module class, i.e., buggy or not buggy). We end up with 17 datasets for our analysis. Table~\ref{tab:a} shows the selected datasets for our study along with basic characteristics about each dataset.

\subsection{Overall approach}

An overview approach of our study is presented in Figure~\ref{fig:framework}. First, we perform correlation analysis and redundancy analysis. Then, we build two defect classifiers, one using the non-discretized defect counts and the other using the discretized defect count (i.e., ``defective'' or ``non-defective''), respectively. After the classifiers are built, we calculate their performance using the \textit{Area Under the receiver operator characteristic Curve (AUC)} and compute the feature importance for each classifier. We repeat this process 1,000 times using out-of-sample bootstrap validation to ensure that our drawn conclusions are statistically robust as suggested by Tantithamthavorn et al.~\cite{40tantithamthavorn2016empirical}. In each iteration of the bootstrap, we compute the AUC values and feature importance for the discretized and regression-based classifiers. We use the computed AUC and feature importances to conduct our preliminary study in Section~\ref{sec:prestudy} and answer our research questions in Section~\ref{sec:Results}.

The individual steps of our approach are explained in detail below.


\subsection{Correlation analysis \& Redundancy analysis}

\textbf{Correlation analysis:} To avoid multicollinearity problems in our classifiers, we perform a correlation analysis to remove highly correlated features. We use a feature clustering analysis to construct a hierarchical overview of the Spearman correlations among  features.
For sub-hierarchies of features with correlations larger than 0.7, we select only one feature from the sub-hierarchy for inclusion into our classifiers. When selecting the feature for inclusion, we select the feature that is simplest to interpret and compute. We use the \textbf{varclus} function from the \textbf{Hmisc} R package in this paper.
For example, Figure~\ref{fig:exp} shows the hierachical clustering of the features of the Poi-3.0 project.
We observe that the features ``wmc'', ``npm'', ``rfc'', and ``loc'' have correlation values larger than 0.7. We choose ``loc'' for inclusion in our classifier as it is relatively easy to compute and explain. We repeat a similar process for other correlated features.

\textbf{Redundancy analysis:} The Correlation analysis handles multicollinearity, but it does not remove redundant features, which are features that do not add additional information with respect to other features~\cite{27yu2004efficient}. The presence of these features distorts the relationship between features and the target variable. Hence, it is important to remove redundant features prior to classifier construction.
We use the \textbf{redun} function from \textbf{rms} R package to remove redundant features. The function drops features iteratively until either no previously constructed model of features achieved an R\textsuperscript{2} above a chosen cutoff threshold (0.9 in our case). Table~\ref{tab:a} shows the number of remaining features in our datasets after employing feature selection on each dataset.


\subsection{Classifier construction}

In our experiments, we use two approaches to build defect classifiers to predict whether a module has defects or not: a traditional defect classifier that is built with discretized defect counts (referred as a \emph{discretized defect classifier}) and a classifier that is built using a regression model that is built with non-discretized defect counts (referred as a \emph{regression-based defect classifier}).

\textbf{Construction of discretized defect classifiers:} The continuous defect counts are discretized to defect classes ``defective'' and ``non-defective'' based on the condition: If a module's defect count is greater than or equal to 1, it is classified as ``defective''; otherwise it is classified as ``non-defective''. During the training phase, the defect classes are then treated as the target variable and are feed to a classification technique (e.g., random forest) along with the collected features to build the discretized classifier. During the testing phase, the trained discretized classifier is tested on unseen testing data to compute the performance (i.e., AUC) of the classifier and its most influential features.

\textbf{Construction of regression-based defect classifiers}: Different from the construction of a discretized classifier, during the training phase, we use the non-discretized defect counts (i.e., the actual values) to build a regression model then the predicted counts of the model are transformed into two classes (``defective'' and ``non-defective'') based on a threshold which is not necessary to be 1. Then the model performance and feature importance are computed.
\begin{figure}
	\includegraphics[width=\linewidth]{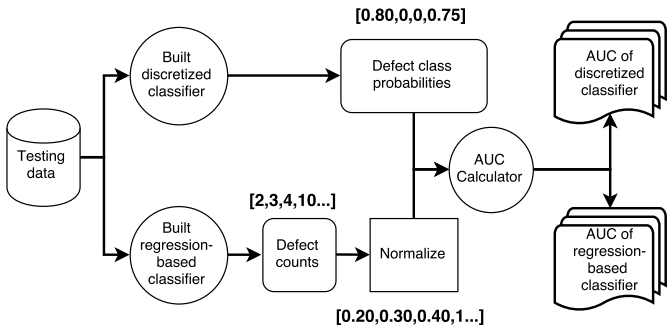}
	\caption{An overview of performance evaluation.}
	
	\label{fig:approach}
\end{figure}

\subsection{Performance calculation}
We use the \textit{Area Under the receiver operator characteristic Curve (AUC)} as the measure of the performance when comparing between the discretized and regression-based defect classifiers~\cite{28lessmann2008benchmarking}. AUC is computed by plotting the ROC curve, which maps the relation between True Positive Rate (TPR) and False Positive Rate (FPR) at all thresholds. We choose AUC because of following reasons:
\begin{enumerate}
  \item AUC measures the performance across all the thresholds. When calculating the performance (e.g., precision and recall) of both discretized and regression-based defect classifiers, a threshold needs to be set up to classify the outcome as ``defective'' if the predicted value is above that threshold and ``non-defective'' otherwise. It is challenging to decide this threshold. To avoid the problem of threshold setting, we select AUC since AUC measures the performance on all the thresholds (i.e., from 0 to 1) and precludes our analysis from the peculiarities of setting up thresholds.
  \item The AUC is insensitive to cost and class distributions~\cite{28lessmann2008benchmarking} so that data imbalances that are inherent to the software datasets is automatically accounted for and provides a score that is objective. An AUC score close to 1 means the classifier's performance is very high and a classifier with an AUC score of 0.5 is no better than random guessing.
\end{enumerate}

We now briefly discuss the calculation of the AUC for the regression-based defect classifier. Figure~\ref{fig:approach} depicts how the performance calculation component of Figure~\ref{fig:framework} works. The AUC calculation is accomplished by normalizing the predicted defect counts to fall within the 0 and 1 range to mimic the class probabilities that are generated by a discretized defect classifier. We use the normalized score along with the actual classes to compute the AUC. In this way, we compare the discretized defect and regression-based defect classifiers on a common ground. This normalized score also ensures that the regression-based defect classifier is tested for its classification prowess rather than its regression performance.

\subsection{Feature importance analysis}
We use permutation feature importance~\cite{29altmann2010permutation} as a means of measuring the importance of a given feature. Understanding the importance of each feature on a classifier, helps practitioners in their process improvement activities for avoiding future defects. We use permutation importance in lieu of the built-in the feature importance algorithm of each classifier and regression technique since permutation importance gives us a way of conducting feature importance estimation in an unbiased setting.

The permutation importance works by randomly permuting the values of one feature at a time so that the original relationship between the feature and target variable is disturbed. Then this permuted feature along with the other non-permuted features is used to classify the testing data, and performance of the classifier is computed. If the computed performance of the classifier that is built using the permuted feature decreases significantly from the classifier that is built with non-permuted feature, then such performance decrease signifies the importance of this feature. This process is repeated for each of the features and they are ranked based on the degree of decrease in the performance once that particular feature is permuted.

\subsection{Out-of-sample bootstrap}

In order to ensure that the conclusions that we draw about our classifiers are robust, we use the out-of-sample bootstrap validation technique, which has been shown to yield the best balance between the bias and variance in a recent study~\cite{40tantithamthavorn2016empirical}. The out-of-sample bootstrap is conducted along the following steps:

\begin{enumerate}
  \item A bootstrap sample of size \textit{N} is randomly drawn with replacement from the original dataset, which is also of size \textit{N}.
  \item Discretized and regression-based defect classifiers are trained using the bootstrap sample (i.e., training data). On average, 36.8\% of the data points will not appear in the bootstrap sample, since it is drawn with replacement~\cite{40tantithamthavorn2016empirical}.
  \item We calculate the AUC value and feature importance for each classifier on unseen testing data that are data points that do not appear in the bootstrap sample.
\end{enumerate}

The out-of-sample bootstrap process is repeated 1,000 times. After the bootstrap, 1,000 AUC values and 1,000 lists of the feature importances are generated. We perform further analysis on these AUC values and feature importance to answer our research questions.




\section{Preliminary Study}
\label{sec:prestudy}

Prior studies have compared the performance of different machine learning classifiers when building discretized defect classifiers~\cite{58guo2004robust,28lessmann2008benchmarking}. However, there is no knowledge about the performance of regression-based classifiers. Hence we focus our preliminary study to examine the performance of regression-based classifiers. Our goals are two folds: 1) To replicate prior findings in order to understand whether prior findings for  discretized defect classifiers would hold for regression-based classifiers, 2) To help focus our analysis in the following sections on the top performing regression-based classifiers.


\textbf{Approach:} We choose one representative classifier from each widely used machine learning families that are listed by previous studies~\cite{36ghotra2015revisiting,28lessmann2008benchmarking} by satisfying following criteria:

\begin{enumerate}
  \item A classifier could be built based on both discretized and non-discretized defect counts.
  \item One is widely used in prior defect prediction studies.
\end{enumerate}


\begin{table}

    \caption{Classifiers that are selected from each family.}\label{tab:classifiers}

    \centering
    \begin{tabular}{p{2.2cm}|p{2.7cm}|p{2.7cm}}
        \hline

        \textbf{Family} & \textbf{Classifier} & \textbf{Regression classifier}  \\

        \hline
        Statistical & Logistic regression (Log-Reg) & Linear regression (Lin-Reg)\\
        \hline
        Random forest & Random forest classification (RF-C) & Random forest regression (RF-R)\\
        \hline
        Neural networks & Neural Networks classifier (NN-C) & Neural Networks regression (NN-R)\\
        \hline
        Decision tree & Classification tree (CT) & Regression tree (RT)\\
        \hline
        Support-Vector machines & SVM classifier (SVM-C) & SVM regression (SVM-R)\\
		\hline
        Nearest neighbor  & K-NN classification (KNN-C) &  K-NN regression (KNN-R)\\
		\hline

    \end{tabular}
\end{table}


Table~\ref{tab:classifiers} shows the classifiers chosen for our analysis. All of these classifiers are used at their default settings. 

We use the overall experiment setup that is outlined in Section~\ref{sec:ExperimentSetup} with all the six families of chosen classifiers. We start with data collection, then correlation and redundancy analysis on the datasets. We then build the discretized and regression-based defect classifiers using the six families of chosen classifiers. The generated classifiers are validated and the performance of each classier is evaluated using the AUC that is obtained from the out-of-sample bootstrap as explained in the Section~\ref{sec:ExperimentSetup}.

Once the AUC values are computed, we use a Scott-Knott Effect size clustering (SK-ESD)~\cite{40tantithamthavorn2016empirical} to rank the classifiers based on the AUC values. SK-ESD uses the effect size as computed by Cohen's $\Delta$~\cite{47cohen1988statistical} to merge statistically similar groups into the same rank. These ranks are obtained for both discretized and regression based defect classifiers on all 17 datasets for each of the six families of chosen classifiers. The average ranks for each classifier over 17 datasets are calculated and the classifier with the lowest rank across both the approaches is considered as the best classifier.

\begin{table*}
\begin{threeparttable}
\caption{Average ranks of various discretized and regression-based  classifiers}\label{tab:RQ1resultsreg}

\centering

\begin{tabular}{l|p{0.2cm}p{0.2cm}|p{0.2cm}p{0.2cm}|p{0.2cm}p{0.35cm}|p{0.2cm}p{0.35cm}|p{0.2cm}p{0.35cm}|p{0.2cm}p{0.35cm}|p{0.2cm}p{0.35cm}|p{0.2cm}p{0.35cm}|p{0.2cm}p{0.35cm}|p{0.2cm}p{0.35cm}|p{0.2cm}p{0.35cm}|p{0.2cm}p{0.2cm}}
    \hline

Project & \multicolumn{2}{c|}{Lin-Reg}   & \multicolumn{2}{c|}{Log-Reg} &  \multicolumn{2}{c|}{RF-R}   & \multicolumn{2}{c|}{RF-C}  & \multicolumn{2}{c|}{NN-R}   & \multicolumn{2}{c|}{NN-C}   & \multicolumn{2}{c|}{RT}   & \multicolumn{2}{c|}{CT}  & \multicolumn{2}{c|}{SVM-R}   & \multicolumn{2}{c|}{SVM-C}  & \multicolumn{2}{c|}{KNN-R}   & \multicolumn{2}{c}{KNN-C} \\
    \hline

  & R & A & R & A & R & A & R & A & R & A & R & A & R & A & R & A & R & A & R & A & R & A & R & A\\
  \hline
Eclipse-2.0  & 2 & 0.80 & 2 & 0.83 & 1 & 0.84 & 1 & 0.84 & 6 & 0.63 & 3 & 0.71 & 5 & 0.71 & 6 & 0.71 & 4 & 0.75 & 5 & 0.76 & 3 & 0.79 & 4 & 0.78\\
Eclipse-2.1 & 1 & 0.79 & 1 & 0.79 & 2 & 0.78 & 2 & 0.77 & 3 & 0.73 & 4 & 0.67 & 3 & 0.73 & 4 & 0.67 & 5 & 0.62 & 5 & 0.62 & 4 & 0.72 & 3 & 0.71\\
Eclipse-3.0  & 1 & 0.80& 1 & 0.80 & 1 & 0.80 & 1 & 0.80 & 5 & 0.62 & 2 & 0.67 & 3 & 0.73 & 4 & 0.67 & 4 & 0.71 & 5 & 0.71 & 2 & 0.75 & 3 & 0.74\\
Camel-1.2 & 3 & 0.60 & 2 & 0.61 & 2 & 0.63 & 1 & 0.64 & 5 & 0.55 & 4 & 0.56 & 3 & 0.55 & 4 & 0.56 & 1 & 0.64 & 5 & 0.64 & 2 & 0.59 & 3 & 0.58\\
Mylyn & 3 & 0.68 & 1 & 0.70 & 1 & 0.74 & 2 & 0.68 & 6 & 0.54 & 2 & 0.60 & 5 & 0.62 & 4 & 0.60 & 4 & 0.65 & 3 & 0.65 & 2 & 0.70 & 1 & 0.70\\
PDE & 2 & 0.69 & 1 & 0.72 & 1 & 0.71 & 2 & 0.71 & 5 & 0.56 & 5 & 0.64 & 4 & 0.64 & 4 & 0.64 & 4 & 0.64 & 6 & 0.65 & 3 & 0.66 & 3 & 0.65\\
Prop-1 & 4 & 0.71 & 2 & 0.74 & 1 & 0.79 & 1 & 0.77 & 5 & 0.62 & 4 & 0.61 & 5 & 0.62 & 4 & 0.61 & 3 & 0.73 & 3 & 0.72 & 2 & 0.76 & 1 & 0.76\\
Prop-2 & 4 & 0.66 & 3 & 0.71 & 1 & 0.84 & 1 & 0.81 & 5 & 0.57 & 5 & 0.50 & 5 & 0.57 & 5 & 0.50 & 3 & 0.68 & 4 & 0.68 & 2 & 0.76 & 2 & 0.75\\
Prop-3 & 3 & 0.68 & 2 & 0.71 & 1 & 0.72 & 4 & 0.69 & 6 & 0.54 & 1 & 0.50 & 5 & 0.58 & 6 & 0.50 & 4 & 0.64 & 5 & 0.64 & 2 & 0.70 & 3 & 0.70\\
Prop-4 & 2 & 0.73 & 1 & 0.75 & 1 & 0.77 & 2 & 0.72 & 5 & 0.63 & 5 & 0.58 & 5 & 0.63 & 5 & 0.58 & 4 & 0.67 & 4 & 0.66 & 3 & 0.71 & 3 & 0.71\\
Prop-5 & 3 & 0.66 & 2 & 0.71 & 1 & 0.73 & 3 & 0.70 & 5 & 0.58 & 1 & 0.51 & 4 & 0.63 & 6 & 0.51 & 3 & 0.66 & 5 & 0.66 & 2 & 0.69 & 4 & 0.69\\
Xalan-2.5 & 5 & 0.63 & 3 & 0.65 & 1 & 0.75 & 1 & 0.76 & 6 & 0.62 & 3 & 0.66 & 4 & 0.64 & 3 & 0.66 & 2 & 0.72 & 4 & 0.72 & 3 & 0.70 & 2 & 0.70\\
Xalan-2.6 & 3 & 0.78 & 2 & 0.80 & 1 & 0.82 & 1 & 0.84 & 5 & 0.66 & 3 & 0.77 & 4 & 0.77 & 3 & 0.77 & 2 & 0.80 & 4 & 0.81 & 3 & 0.79 & 2 & 0.81\\
Lucene-2.4 & 2 & 0.75 & 2 & 0.74 & 1 & 0.77 & 1 & 0.77 & 5 & 0.50 & 5 & 0.67 & 4 & 0.67 & 4 & 0.67 & 3 & 0.72 & 6 & 0.73 & 3 & 0.72 & 3 & 0.71\\
Poi-2.5 & 5 & 0.76 & 3 & 0.81 & 2 & 0.85 & 1 & 0.89 & 6 & 0.50 & 3 & 0.80 & 4 & 0.79 & 3 & 0.80 & 1 & 0.86 & 4 & 0.86 & 3 & 0.82 & 2 & 0.84\\
Poi-3.0 & 5 & 0.75 & 2 & 0.84 & 1 & 0.82 & 1 & 0.89 & 6 & 0.50 & 4 & 0.82 & 4 & 0.75 & 3 & 0.82 & 3 & 0.80 & 5 & 0.85 & 2 & 0.81 & 2 & 0.85\\
Xerces-1.4 & 5 & 0.86 & 1 & 0.94 & 1 & 0.91 & 1 & 0.96 & 6 & 0.50 & 4 & 0.91 & 5 & 0.86 & 3 & 0.91 & 3 & 0.90 & 5 & 0.91 & 2 & 0.90 & 2 & 0.92\\

\hline
\textbf{Avg.} & 3.12 & 0.73 & 1.82 & 0.76 & \textbf{1.17 }& \textbf{0.78} & \textbf{1.52} & \textbf{0.78} & 5.29 & 0.58 & 3.41 & 0.66 & 4.23 & 0.68 & 4.18 & 0.66 & 3.11 & 0.72 & 4.59 & 0.72 & 2.52 & 0.74 & 2.52 & 0.74\\
\hline

\end{tabular}
\begin{tablenotes}
    \small
    \item R - Rank; A - AUC
    
\end{tablenotes}
\end{threeparttable}
\end{table*}

\textbf{Results: } \textbf{The random forest family has the best performance across both discretized and regression-based defect classifiers.} The ranks for each classifier are listed in Table~\ref{tab:RQ1resultsreg}. Random forest family has the best performance across both discretized (i.e., average rank is 1.17) and regression-based defect classifiers (i.e., average rank is 1.52). This is compatible with the findings of prior studies that suggest that a random forest classifier outperforms other classifiers for building discretized defect classifiers~\cite{36ghotra2015revisiting,28lessmann2008benchmarking}. More specifically, the random forest family performs the best in 14 out of 17 studied datasets for regression-based defect classification and in 11 out of 17 studied dataset for discretized defect classification.

The next best classifier is K-Nearest Neighbor family which has an average rank of 2.52 for both the regression-based and discretized defect classifier, followed by the Statistical classifiers (i.e., Linear and Logistic regressions) which have an average rank of 3.12 for discretized defect classifiers and 1.82 for the regression-based defect classifiers. 

We also report the average AUC for each classifier on the studied datasets in Table~\ref{tab:RQ1resultsreg}. We find that random forest family has an average AUC of 0.78 across both discretized and regression-based defect classifiers, which is the highest among all considered classifiers. The average AUC for linear and logistic regression is 0.73 for regression-based defect classifiers and 0.76 for discretized defect classifiers, which is followed by the KNN family at 0.74 for both types of classifiers.

\rqbox{\textit{The random forest family performs the best across both discretized and regression-based defect classifiers. Hence, we primarily focus our analysis in the following sections on the random forest family.}}

\section{Case Study Results}
\label{sec:Results}

\subsection{\textbf{RQ1. How well do regression-based classifiers perform?}}

\textbf{Motivation:} In this research question, we investigate the performance of regression-based defect classifiers. Both regression-based and discretized classifiers can identify defect-prone modules. Prior research in software engineering has primarily used discretized classifiers for identifying defect-prone modules. However the discretization that is performed by discretized classifiers leads to information loss. Hence, regression-based classifiers might be a viable option. 

\textbf{Approach:} To answer this research question, we construct discretized and regression-based classifiers on the 17 studied datasets. Based on our observations in Section~\ref{sec:prestudy}, we use random forest classifiers since they outperform other types of classifiers.

We then compare the performance of the discretized random forest classifiers (DRFC) and regression-based random forest classifiers (RBRFC) using the AUC values. To measure the differences between the two types of classifiers, we use a Wilcoxon signed-rank test~\cite{46wilcoxon1945individual} since it does not need the data to follow a normal distribution and it tests paired results. To quantify the magnitude of the performance differences between DRFC and RBRFC, we use Cohen's $d$ effect size test~\cite{47cohen1988statistical}. The threshold for analyzing the magnitude is as follows: $|d|$ $\leq$ 0.2 means magnitude is negligible,  $|d|$ $\leq 0.5$ means small, $|d|$ $\leq 0.8$ means medium and $|d|$ $>$ 0.8 means large.


\begin{table}
    \begin{threeparttable}

        \caption{Performance Comparison of discretized and regression-based random forest classifiers.}\label{tab:RQ2results}

        \centering
        \begin{tabular}{l|p{1.5cm}p{1.5cm}p{0.6cm}p{1.2cm}p{0.5cm}}
            \hline

\textbf{Project} & \textbf{Avg. AUC of DRFC} & \textbf{Avg. AUC of RBRFC} & \textbf{$p$-Value}  & \textbf{Cohen's $d$}& \textbf{DR(\%)} \\
\hline

Prop-4 & \multicolumn{1}{r}{0.72} & \multicolumn{1}{r}{\textbf{0.77}} & \multicolumn{1}{r}{0}& \multicolumn{1}{r}{4.27 (L)} & \multicolumn{1}{r}{9.6}\\
Prop-2 & \multicolumn{1}{r}{0.81} & \multicolumn{1}{r}{\textbf{0.84}} &  \multicolumn{1}{r}{0}  & \multicolumn{1}{r}{4.32 (L)} & \multicolumn{1}{r}{10.5}\\
Eclipse-2.1 & \multicolumn{1}{r}{0.77} & \multicolumn{1}{r}{\textbf{0.78}} & \multicolumn{1}{r}{0}& \multicolumn{1}{r}{0.46 (S)} & \multicolumn{1}{r}{10.8}\\
Prop-3 & \multicolumn{1}{r}{0.69}& \multicolumn{1}{r}{\textbf{0.72}} & \multicolumn{1}{r}{0}& \multicolumn{1}{r}{2.67 (L)} & \multicolumn{1}{r}{11.5}\\
Mylyn & \multicolumn{1}{r}{0.68} & \multicolumn{1}{r}{\textbf{0.74}} & \multicolumn{1}{r}{0}& \multicolumn{1}{r}{2.45 (L)} & \multicolumn{1}{r}{13.2}\\
PDE & \multicolumn{1}{r}{0.71} & \multicolumn{1}{r}{0.71} & \multicolumn{1}{r}{0}& \multicolumn{1}{r}{0.12 (N)}  & \multicolumn{1}{r}{14.0}\\
Eclipse-2.0 & \multicolumn{1}{r}{0.84} & \multicolumn{1}{r}{0.84} & \multicolumn{1}{r}{0}& \multicolumn{1}{r}{0.22 (S)} & \multicolumn{1}{r}{14.5}\\
Eclipse-3.0 & \multicolumn{1}{r}{0.80} & \multicolumn{1}{r}{0.80} & \multicolumn{1}{r}{0.04} & \multicolumn{1}{r}{-0.03 (N)} & \multicolumn{1}{r}{14.8}\\
Prop-1 & \multicolumn{1}{r}{0.77} & \multicolumn{1}{r}{\textbf{0.79}} & \multicolumn{1}{r}{0}& \multicolumn{1}{r}{2.93 (L)} & \multicolumn{1}{r}{14.8}\\
Prop-5 & \multicolumn{1}{r}{0.70} & \multicolumn{1}{r}{\textbf{0.73}} & \multicolumn{1}{r}{0}& \multicolumn{1}{r}{2.97 (L)} & \multicolumn{1}{r}{15.3}\\
Camel-1.2 & \multicolumn{1}{r}{\textbf{0.64}} & \multicolumn{1}{r}{0.63} & \multicolumn{1}{r}{0}& \multicolumn{1}{r}{-0.39 (S)} & \multicolumn{1}{r}{35.5}\\
Xalan-2.6 & \multicolumn{1}{r}{\textbf{0.84}} & \multicolumn{1}{r}{0.82} & \multicolumn{1}{r}{0}& \multicolumn{1}{r}{-0.96 (L)} & \multicolumn{1}{r}{46.4}\\
Xalan-2.5 & \multicolumn{1}{r}{\textbf{0.76 }}& \multicolumn{1}{r}{0.75} & \multicolumn{1}{r}{0}& \multicolumn{1}{r}{-0.39 (S)} & \multicolumn{1}{r}{48.2}\\
Lucene 2.4 & \multicolumn{1}{r}{0.77} & \multicolumn{1}{r}{0.77} & \multicolumn{1}{r}{0}& \multicolumn{1}{r}{-0.16 (N)} & \multicolumn{1}{r}{59.7}\\
Poi 3.0 & \multicolumn{1}{r}{\textbf{0.89}} & \multicolumn{1}{r}{0.82} & \multicolumn{1}{r}{0}& \multicolumn{1}{r}{-2.42 (L)} & \multicolumn{1}{r}{63.6}\\
Poi 2.5 & \multicolumn{1}{r}{\textbf{0.89}} & \multicolumn{1}{r}{0.85} & \multicolumn{1}{r}{0}& \multicolumn{1}{r}{-1.46 (L)} & \multicolumn{1}{r}{64.4}\\
Xerces 1.4 & \multicolumn{1}{r}{\textbf{0.96}} & \multicolumn{1}{r}{0.91} & \multicolumn{1}{r}{0}& \multicolumn{1}{r}{-2.43 (L)} & \multicolumn{1}{r}{74.3}\\

            \hline
        \end{tabular}

        \begin{tablenotes}
            \small
            \item L- Large, S- Small, N- Negligible, DR - Defective Ratio
        \end{tablenotes}
    \end{threeparttable}
\end{table}

\textbf{Results:} \textbf{In contrast to prior studies, building a defect classifier using discretized defect counts does not usually lead to better performance.} The comparison of DRFC and RBRFC of all datasets are provided in Table~\ref{tab:RQ2results}. Overall, the Wilcoxon signed-rank test results show that the differences between DRFC and RBRFC are significant on all datasets. Cohen's $d$ results show that the performance differences between DRFC and RBRFC are not negligible for 14 datasets (82\%). More specifically, on 7 out of these 14 datasets, DRFC outperforms RBRFC; while, in contradiction to the intuition, RBRFC outperforms DRFC on another 7 datasets.

\begin{figure}
	
	\includegraphics[width=\linewidth]{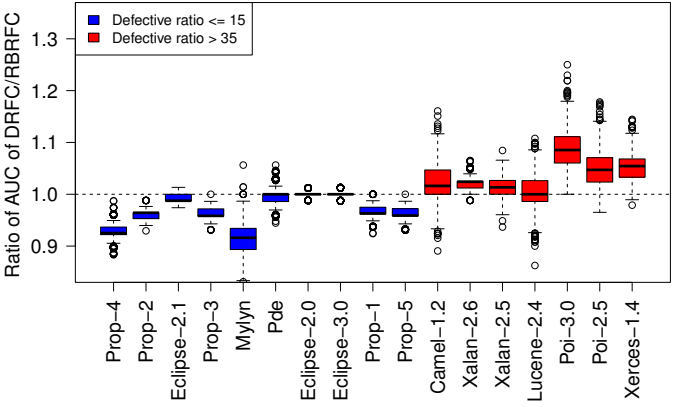}
	\caption{Ratio of AUC of discretized/regression-based random forest classifiers across different datasets. The datasets are ordered based on their defective ratio from low to high.}
	
	\label{fig:fig2}
\end{figure}

\textbf{Regression-based random forest classifiers outperform discretized random forest classifiers when the defective ratio of the dataset data is less than 15\% and this trend is reversed when the defective ratio is greater than 35\%}. To understand how the performance varies among different datasets, we plot the ratio of AUCs of DRFC and RBRFC on the y-axis of Figure~\ref{fig:fig2} and we sort the datasets from low defective ratio to high defective ratio on the x-axis. We observe a consistent trend of RBRFC outperforming DRFC for datasets with a low defective ratio ($< 15\%$) and DRFC outperforms RBRFC when the defective ratio is greater than 35\%. It should also be noted that though the difference in average AUC is only a few percentage points, it is statistically significant as highlighted in Table~\ref{tab:RQ2results}.

One possible reason for DRFC having poorer performance than RBRFC on datasets with low defective ratios is that discretized random forest classifiers are known to be impacted by the imbalance in the dataset~\cite{63chen2004using}. It is the fact in our case, findings are suggestive of the fact that, these imbalanced datasets can be better handled by using a regression-based random forest defect classifier in lieu of the traditionally-used discretized classifiers.

\begin{figure}
	\includegraphics[width=\linewidth]{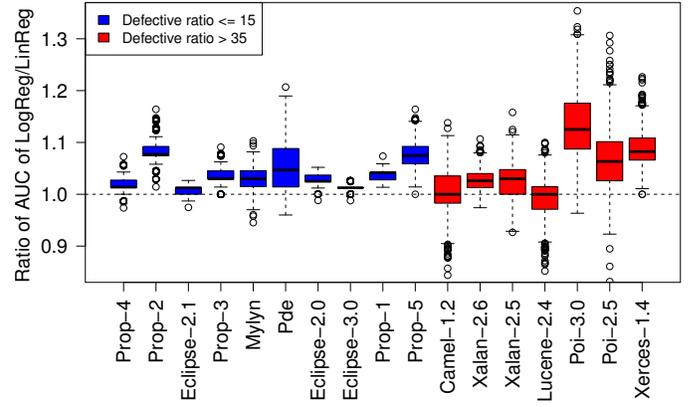}
	\caption{Ratio of AUC of discretized/regression-based statistical defect classifiers across different datasets. The datasets are ordered based on their defective ratio from low to high.}
	
	\label{fig:fig3}
\end{figure}

\textbf{The trend of discretized classifiers outperforming regression-based defect classifiers is unique to random forest classifiers.} No similar trend is observed for other types of classifiers (e.g., LogReg, LinReg, and KNN). When using other types of classifiers, the discretized  classifiers either outperform or perform as good as the regression-based  classifiers. For example, Figure~\ref{fig:fig3} shows the ratio of AUCs between discretized and regression-based statistical classifiers. We  observe that the medians of the ratios are all above the dashed line which indicates that discretized logistic classifiers  always outperform or perform similar toh regression-based linear classifier. More detailed results are available in Table~\ref{tab:RQ1resultsreg}.

In summary, the common intuition of building a classifier using discretized defect counts is not always correct. To achieve high performance, we advise the use of RBRFC instead of its discretized alternative on datasets with a low defective ratio (i.e., less than 15\%) and the use of DRFC on datasets with a high defective ratio. For datasets with a defective ratio between 15\% and 35\%, we cannot provide suggestions on which classifier to use, since we do not have datasets in that range of defective ratio. To alleviate this problem, we revisit this point in Section~\ref{sec:Discussion} where we simulate datasets with different defective ratios.

\rqbox{\textit{In constrast to the common practice, building a defect classifier using discretized defect counts does not always lead to better performance. RBRFC outperform DRFC when the defective ratio of the dataset is low ($< 15\%$) and the pattern reverses when the defective ratio is high ($> 35\%$).}}

\subsection{\textbf{RQ2. Are discretized and regression-based classifiers influenced by the same set of features?}}

\textbf{Motivation:} Prior studies use defect classifiers to understand the impact of various features (e.g., software metrics) on the likelihood of a module containing a defect~\cite{12cataldo2009software,13mcintosh2014impact,14mockus2010organizational}. Understanding the most influential features helps practitioners identify process improvement plans and act on them quickly so that the defects could be avoided in future version of a software system. In this paper, we introduce an approach to build defect classifiers using regression. In RQ1, we find that such an approach might lead to better performing classifiers than the traditionally used approach for building classifiers (i.e., discretized classifiers). Hence, in this RQ we wish to examine if these different approaches to build classifiers might produce conflicting information about the most influential features that impact the quality of a software module.

\textbf{Approach:} Similar to the previous research question, we focus on exploring this RQ with DRFC and RBRFC. However, we will provide insights about the other machine learning classifiers whenever appropriate. We follow the approach in Figure~\ref{fig:framework}. We build DRFC and RBRFC on the 17 studied datasets as in RQ1. But instead of generating the AUC values of the classifiers, we generate the feature importance for each feature in each dataset to understand the importance of each feature in identifying the defect-proneness of modules. We use a permutation importance method for generating feature importance scores for both DRFC and RBRFC as outlined in section~\ref{sec:ExperimentSetup}. Once the feature importance scores are generated, we rank features using the Scott-Knott ESD test~\cite{40tantithamthavorn2016empirical}. We then compare the feature importance ranks of DRFC and RBRFC.



To estimate the impact of DRFC and RBRFC on model interpretation, we compute the shifts in the ranks of the features that appear in the top three ranks for both the DRFC and RBRFC classifiers on each dataset. We define rank shift as the amount that a feature shifts its rank between the two classifier in relation to the total number of features in the dataset. Suppose $DRFC(k) = \{var_1,var_2,...,var_n\}$ and $RBRFC(k) = \{var_1,var_2,...,var_m\}$ are the features that appear at rank $k$ of $DRFC$ and $RBRFC$. Let $PN$ be the number of features in the given dataset under consideration. We compute the $Shifts(k)$ of a features on rank $k$ between two classifier for a given dataset using the equation~(\ref{eqn1}).

\begin{equation}
\begin{aligned}
 Shifts(k) =
(\sum_{var \in DRFC(k)}^{}|k-Rank_{RBRFC}(var)|\\
+ \sum_{var \in RBRFC(k)}^{}|k-Rank_{DRFC}(var)| )/PN\\
\end{aligned}\label{eqn1}
\end{equation}
where $Rank_{RBRFC}(var)$ denotes that rank of $var$ from RBRFC and $Rank_{DRFC}(var)$ denotes the rank of $var$ from DRFC. For example, if the Rank 1 features in the RBRFC are $RBRFC(1)=\{cbo,loc\}$ (i.e., Coupling Between Objects and Lines Of Code), Rank 1 features for $DRFC$ is $DRFC(1)=\{loc\}$ and $DRFC(2)=\{cbo\}$, hence $Rank_{DRFC}(cbo)$ is 2 and $NP$ is 13 for the dataset under consideration, we then compute the $Shifts(1)$ between both classifiers as $1/13 = 0.076$, since only the feature $cbo$ has different ranks across both classifiers. We compute the feature importance shifts for all datasets in a similar fashion. These rank shifts capture the difference in the influential features across the two approaches for building classifiers.


\begin{table}
    \caption{Rank shifts between discretized and regression-based classifiers on various classifiers.}\label{tab:averages}

    \centering
    \begin{tabular}{p{2cm}|l|r|r}
        \hline
        \textbf{Technique} & \textbf{Rank} & \textbf{Average Shifts} & \textbf{Variance}  \\
        \hline
        & Rank 1 & 0.04 & 0.004 \\
 Random forest&Rank 2 & 0.07 & 0.007 \\
&Rank 3 & 0.16 & 0.03\\

        \hline

        & Rank 1 & 0.11 & 0.22 \\
        Statistical &Rank 2 & 0.07 & 0.005 \\
        &Rank 3 & 0.22 & 0.039\\

        \hline
        & Rank 1 & 0.01 & 0.001 \\
        KNN  &Rank 2 & 0.02 & 0.002 \\
        &Rank 3 & 0.07 & 0.008\\

        \hline
    \end{tabular}
\end{table}

\begin{figure}
    \includegraphics[width=\linewidth]{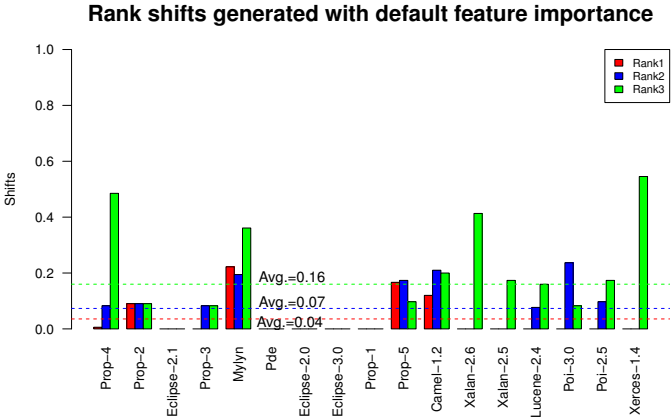}
    \caption{Rank shifts between DRFC and RBRFC in terms of permutation feature importance  across the datasets ordered by defective ratio of the dataset. The mean values of rank shifts are marked with dashed lines.}
    
    \label{fig4:difference}
\end{figure}

\textbf{Results: }\textbf{Rank 1 features do not vary significantly between the DRFC and RBRFC classifiers, however the influential features vary significantly at the lower ranks.} Figure~\ref{fig4:difference} shows the rank differences between he DRFC and RBRFC classifiers. The DRFC and RBRFC classifiers have exactly the same rank 1 features  in 12 out of the 17 datasets (80\%). The features at rank 2 and 3 vary drastically since only 8 (47\%) and 5 (29\%) datasets have the same features at rank 2 and 3, respectively. In terms of rank shift, we observe that the shift between features in rank 1 (i.e., average shift is 0.04) is small and the feature importance varies slightly at rank 2 (i.e., average shift is 0.07). But from rank 3, the shifts start becoming drastic (i.e., 0.16). The dashed horizontal lines that represent the average shift in each rank between the features in Figure~\ref{fig4:difference}). We also performed a paired Wilcoxon signed-rank test between the observed shifts and ideal no shift case in which each shifts value is 0 for all datasets. The results show that the shifts at rank 1 are not statistically significant ($p$-value $>$ 0.05) and shifts at rank 2 and 3 are significant.

We also investigate the rank shifts between discretized and regression-based classifiers for families other than random forest. We present the findings of statistical and KNN classifiers in Table~\ref{tab:averages} as they are the next best classifiers after random forests in terms of performance. We find that KNN classifiers exhibit a similar pattern as the random forest classifiers. The feature importance of Rank 1 features does not vary significantly, nevertheless the feature importances start to vary significantly from Rank 2. However, when using statistical classifiers the importance of features shift significantly even at Rank 1.

In summary, although the rank shifts of features appear to be family dependent, random forest has the most stable ranks for its features across both approaches for building defect classifiers. Nevertheless, we recommend that feature importance of the best performing classifier should be used instead of relying solely on the feature importance results that are produced by the discretized classifiers (since such classifiers might fail to accurately capture the studied datasets as observed in some case in RQ1).

\rqbox{The importance of Rank1 features does not vary significantly between discretized and regression-based random forest classifiers. However, lower ranked features vary considerably between types of classifiers. Thus, we suggest practitioners to employ caution on the feature importance variation and use the classifiers with superior performance for model interpretation.}

\section{Discussion}

\label{sec:Discussion}
This discussion section is evaluated in the context of random forest based discretized and regression-based defect classifiers.
\subsection{How does the performance of discretized and regression-based random forest classifiers vary across different defective ratio?}
In RQ1, we observe that the RBRFC outperforms DRFC on data with defective ratio larger than 35\% and the observations reverses on the data with defective ratio less than 15\%. However, we do not know how the performance of DRFC and RBRFC varies on the data with defective ratio between 15\% and 35\%. To fill this gap and better understand the relation between the defective ratio and the performance difference between both approaches for building defect classifiers, we generate datasets with defective ratio ranging from 5\% to 50\% with 5\% interval by re-sampling the studied datasets with replacement (i.e. while keeping the data size fixed). We do this by repeatedly and randomly sampling the datasets with replacement until the datasets have the required defective ratio for our simulation study. Once the datasets at all defective ratios are generated,
we followed the experiment setup of RQ1 and analyzed the performance of the generated DRFC and RBRFC for various defective ratios.

We find that as a dataset's defective ratio increases the DRFC classifiers start to outperform the RBRFC classifiers.
Table~\ref{tab:correlations} shows the spearman correlation ($\rho$) between the ratio of
$AUCofDRFC(dataset)/AUCofRBRFC(dataset)$ and the defective ratio of each dataset.
For most datasets (94\% -- 16 out of 17 datasets ), there is indeed a positive and strong  correlation (i.e., $>0.5$) between the defective
ratio and the ratio of the AUC in 16 out of 17 datasets.

\begin{figure}
	\includegraphics[width=\linewidth]{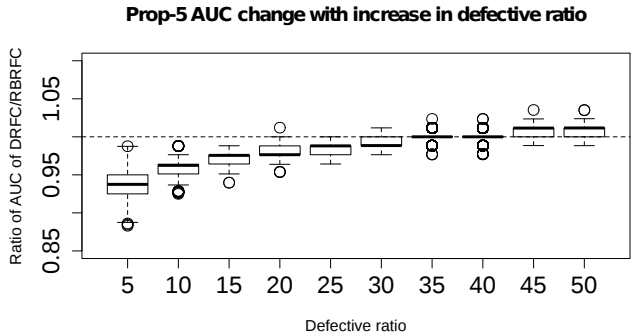}
	
	\caption{Boxplot of the ratio of AUC of DRFC/RBRFC on Prop-5.}
	
	\label{fig4:disc1}
\end{figure}

For example, we show how the DRFC outperforms RBRFC as the defective ratio of the dataset increases using the ``Prop-5'' dataset in Figure~\ref{fig4:disc1}. After the defective ratio crosses 40\%, the DRFC outperforms the RBRFC. This study of variation in performance of AUC between DRFC and RBRFC reaffirms our
findings in the RQ1 that RBRFC perform better for
datasets with low defective ratio whereas the DRFC perform
better as the defective ratio of the dataset increases.
\begin{table}
	
	\caption{Correlation between defective ratio and ratio of AUC of DRFC/RBRFC.}\label{tab:correlations}
	
	\centering
	\begin{tabular}{l|r|l|r}
		\hline
		\textbf{Dataset} & \textbf{Correlation}   & \textbf{Dataset} & \textbf{Correlation} \\
		\hline
		Eclipse-2.0 & 0.90 &Eclipse-2.1 & 0.81\\
		Eclipse-3.0 & 0.84 & Camel-1.2 & 0.89 \\
		Mylyn & 0.79 & Pde & 0.78 \\
		Prop-1 & 0.90 &	Prop-2 & 0.88 \\
		Prop-3 & 0.92 & Prop-4 & 0.91 \\
		Prop-5 & 0.95 & Xalan-2.5 & 0.39 \\
		Xalan-2.6 & 0.86 & 	Lucene-2.4 & 0.51 \\
		Poi-2.5 & 0.78 & Poi-3.0 & 0.56\\
		Xerces-1.4 & 0.54 & & \\
		\hline

	\end{tabular}

\end{table}

Also, we find that the specific point
where DRFC start outperforming RBRFC is dataset specific. But as a general rule of thumb,
DRFC outperform RBRFC as the defective ratio in the
dataset increases. Finally, for datasets with defective ratio between 15\% and 35\% we suggest practitioners to try both DRFC and RBRFC and use the classifier with superior performance.

\subsection{Does the $R^2$ regression fit score impact the performance of regression-based classifiers?}
The $R^2$ regression score explains the variability in the data that is captured by the regression model. Prior studies consider that a higher $R^2$ is usually associated with better performance and more accurate model interpretation~\cite{54nagappan2006mining}. But as we are using the regression-based classifier for the purposes of classification (regression-based classifier), we find that this assumption no longer holds.

\textbf{Low $R^2$ scores for the regression model does not imply that a regression-based classifier will have a low AUC.} There is no correlation between $R^2$ and a classifier's predictive power (i.e., AUC). A low $R^2$ score does not indicate poor classification performance. Table~\ref{tab:r2rmse} presents the values of $R^2$ and the correlation between the AUC and $R^2$ of the RBRFC classifier. Overall, the average AUC across all dataset is good (i.e., 0.78), while the average $R^2$ is poor (i.e., 0.19). We find that in most of the datasets (88\%), the correlation between AUC and $R^2$ is considered weak (i.e., less than 0.4)~\cite{StatisticsInANutshell}. Only two datasets (i.e., Prop-1 and Xerces-1.4), the correlation between AUC and $R^2$ is considered as moderate~\cite{StatisticsInANutshell}. For example, the RBRFC classifier achieves a high AUC 0.84 on Prop-2, while its $R^2$ is only 0.12 and the correlation between the AUC and the $R^2$ is 0.17. 




\begin{table}
	
	\caption{Correlation between AUC and $R^2$ for the RBRFC classifiers. }\label{tab:r2rmse}
	
	\centering
	\begin{tabular}{l|R{1cm}R{0.9cm}|l|R{1cm}R{0.9cm}}
		\hline
		\textbf{Dataset} & \textbf{Correlation} & \textbf{Avg. $R^2$} & \textbf{Dataset} & \textbf{Correlation} & \textbf{Avg. $R^2$} \\
		\hline
Eclipse-2.0 & 0.14 & 0.29 & Eclipse-2.1 & 0.33 & 0.19\\
Eclipse-3.0 & 0.19 & 0.31 & Camel-1.2 & 0.32 & 0.04 \\
Mylyn & 0.28 & 0.05 & Pde & 0.21 & 0.02\\
Prop-1 & 0.08 & 0.05 & Prop-2 & 0.17 & 0.12\\
Prop-3 & 0.21 & -0.05 & Prop-4 & 0.18 & 0.11 \\
Prop-5 & 0.21 & 0.10 & Xalan-2.5 & 0.41 & 0.16\\
Xalan-2.6 & 0.45 & 0.34 & Lucene-2.4 & 0.13 & 0.33 \\
Poi-2.5 & 0.37 & 0.41 & Poi-3.0 & 0.11 & 0.32 \\
Xerces-1.4 & 0.07 & 0.44 &&& \\

		\hline
	\end{tabular}
\end{table}


\subsection{Permutation feature importance vs. Default feature importance?}
In RQ2, we propose permutation feature importance as the method of choice for generating feature importance scores. However, researchers primarily use the default feature importance method that comes along with the implementation of their classifiers. We examine here whether our findings for RQ2 would hold when the default feature importance method is used. We use random forest classifiers for our investigation here. However, the observations hold for all the studied family of classifiers.



\begin{figure}
	\includegraphics[width=\linewidth]{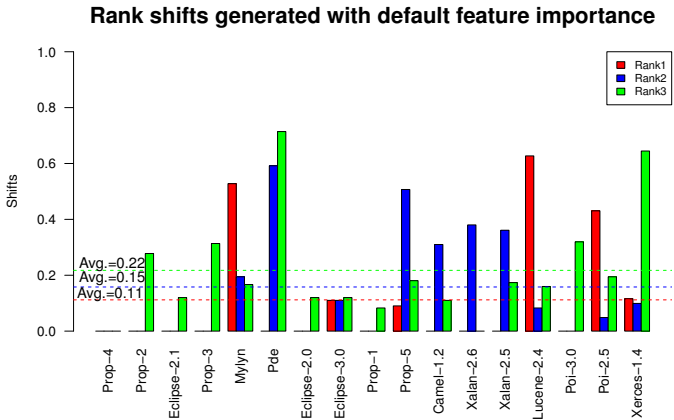}
	\caption{Rank shifts between DRFC and RBRFC in terms of default feature importance across the datasets. The mean values of rank shifts are marked with dashed lines.}
	
	\label{fig7:disc3}
\end{figure}

Figure~\ref{fig4:difference} and~\ref{fig7:disc3} show the rank shifts between permutation and default feature importance methods. We observe that the rank shifts of Rank 1 between DRFC and RBRFC are low (i.e., average shift is 0.11) with default feature importance. However, compared with the rank shifts from permutation feature importance as shown in Figure~\ref{fig4:difference}, the default feature importance has a higher average rank shift at rank 1. We also conduct the Wilcoxon signed-rank test and it suggests that the feature importances that are computed with default method vary significantly from rank 1. The findings that are observed from permutation and default feature importance methods are still hold from rank 2, although the findings are different at rank 1.


\section{Threats To Validity}
\label{sec:ThreatsToValidity}
We discuss the threats to the validity of our study.

\textbf{Construct Validity.} Threats to construct validity relates to the suitability of our evaluation measures. We have used AUC to evaluate the performance of defect classifiers in our study. While we have explained our reasons for choosing this metric,
other evaluation measures may lead different conclusions. For instance, we need to evaluate recall after reading 20\% of the lines of code~\cite{Kamei:2010:RCB:1912607.1913298} but we present it as an avenue of future research. However, AUC is a well-known metric to evaluate classification models and also widely used in prior studies~\cite{28lessmann2008benchmarking,40tantithamthavorn2016empirical}. We have also performed statistical tests and effect size tests to check if the performance differences between different classifiers are significant and substantial.

In this study, we did not optimize the parameters for the studied classifiers except Neural networks, and CART, as most studied classifiers do not get a significant performance boost with parameters optimization ~\cite{37tantithamthavorn2016automated}. However, to reduce this threats, future studies should examine the impact of optimized parameters on our findings.

\textbf{Internal Validity.} Prior work shows that incorrect data influences the conclusions drawn from software defect classifiers and
potentially biases the results~\cite{36ghotra2015revisiting}. Even though we tried to control the purity of datasets by imposing conditions
on data collection, we cannot ensure that our datasets are correct. To reduce the internal validity, future studies should investigate the correctness of the data further.

\textbf{External Validity.}
Threats to external validity relate to the generalizability of our results. In this study, we study 17 datasets and our results may not generalize to other datasets.
However,
the goal of this paper is not to show a result that
generalizes to all datasets, but rather to show that there are
datasets where regression-based classifiers would outperform the commonly used discretized classifiers. Nonetheless,
additional replication studies may prove fruitful.


\section{Conclusion}
\label{sec:Conclusion}
Defect classifiers support software quality assurance efforts in identifying defect-prone modules and allocating quality improvements resources in a timely and cost effective fashion. Traditionally, software defect classifiers are built by discretizing the continuous defect counts of modules into ``defective'' and ``non-defective'' classes. However discretization of continuous variables leads to a considerable loss of information. To avoid such information loss, we consider a regression-based classifiers which use the continuous defect counts as the target variable for identifying defect-prone modules.

In this study, we compare discretized and regression-based defect classifiers by applying six machine learning classifiers on 17 open datasets from Tera-PROMISE. We observe that
in contrast to current practices in our field, building classifiers using discretized defect counts does not always lead to better performance. Hence future studies should explore both approaches for building classifiers -- Given the simplicity of building both types of classifiers, we believe that our suggestion is
a rather simple and low-cost suggestion to follow. Moreover, the most influential features vary between the different approaches to build classifiers. Hence future studies should examine the influential factors using the best performing classifier (i.e., discretized or regression-based) instead of simply using discretized classifiers.




\section{Acknowledgment}
\label{sec: Acknowledgment}
This study would not have been possible without the High Performance Computing (HPC) systems provided by the Compute Canada and Centre for Advanced Computing (CAC). The research was partially supported by JSPS KAKENHI Grant numbers 15H05306.

\balance
\bibliographystyle{abbrv}
\bibliography{main}

\end{document}